\documentclass[preprint]{aastex}

\slugcomment{KSUPT-01/2 \hspace{0.5truecm} February 2001}

\def\uK{\mu {\rm K}}
\def\deg{^{\circ}}

\def\fdg{\hbox{$.\!\!^\circ$}}

\tighten
 
\begin{document}

\title{Gaussianity of Degree-Scale Cosmic Microwave Background Anisotropy \\ 
       Observations}
\author{Chan-Gyung Park\altaffilmark{1}, 
        Changbom Park\altaffilmark{1},
        Bharat Ratra\altaffilmark{2}, 
        and 
        Max Tegmark\altaffilmark{3}}

\altaffiltext{1}{Department of Astronomy, Seoul National University, 151-742   
                 Korea; parkc@astro.snu.ac.kr, cbp@astro.snu.ac.kr.}
\altaffiltext{2}{Department of Physics, Kansas State University, Manhattan, 
                 KS 66506; ratra@phys.ksu.edu.}
\altaffiltext{3}{Department of Physics, University of Pennsylvania, 
                 Philadelphia, PA 19104; max@physics.upenn.edu.}

\begin{abstract}
We present results from a first test of the Gaussianity of degree-scale cosmic 
microwave background (CMB) anisotropy. We investigate Gaussianity of the CMB 
anisotropy by studying the topology of CMB anisotropy maps from the QMAP and 
Saskatoon experiments. We also study the QMASK map, a combination map of the 
QMAP and Saskatoon data. We measure the genus from noise-suppressed 
Wiener-filtered maps at an angular scale of about $1\fdg5$. To test the 
Gaussianity of the observed anisotropy, we compare these results to those 
derived from a collection of simulated maps for each experiment in a Gaussian 
spatially-flat cosmological constant dominated cold dark matter model. The 
genus-threshold level relations of the QMAP and Saskatoon maps are consistent 
with Gaussianity. While the combination QMASK map has a mildly non-Gaussian 
genus curve which is not a consequence of known foreground contamination, this 
result is not statistically significant at the 2 $\sigma$ level. These results 
extend previous upper limits on the non-Gaussianity of the large angular scale 
($> 10\deg$) CMB anisotropy (measured by the COBE DMR experiment) down to degree
angular scales.
\end{abstract}


\section{Introduction}

An important feature of the simplest inflation models is that the initial 
density fluctuation field has a Gaussian random phase distribution (see, e.g., 
Fischler, Ratra, \& Susskind 1985). Therefore, an observational test of the 
Gaussianity of the initial density fluctuation field will provide an important 
constraint on inflation models. Since the CMB temperature anisotropy fluctuation reflects the density distribution on the last scattering surface, one can study 
the Gaussianity of the primordial density fluctuation field by measuring the topology of the CMB anisotropy. 

Topology measures for two-dimensional fields, introduced by Coles \& Barrow 
(1987), Melott et al. (1989), and Gott et al. (1990), have been applied to many 
different kinds of observational data. These include the angular distribution 
of galaxies on the sky (Coles \& Plionis 1991; Gott et al. 1992), the 
distribution of galaxies in slices of the universe (Park et al. 1992; Colley 
1997; Park, Gott, \& Choi 2001), and the CMB temperature anisotropy field 
(Coles 1988; Gott et al. 1990; Smoot et al. 1994; Kogut et al. 1996; Colley, 
Gott, \& Park 1996). 

Recently, there have been many analyses of the Gaussianity of the CMB 
anisotropy at large ($>10\deg$) angular scales that make use of the DMR maps 
(Colley et al. 1996; Kogut et al. 1996; Ferreira, Magueijo, \& G\'orski 1998; 
Novikov, Feldman, \& Shandarin 1999; Pando, Valls-Gabaud, \& Fang 1998;
Bromley \& Tegmark 1999; Banday, Zaroubi, \& G\'orski 2000; Magueijo 2000;
Mukherjee, Hobson, \& Lasenby 2000; Barreiro et al. 2000; Aghanim, Forni, 
\& Bouchet 2001; Phillips \& Kogut 2001). While it has been argued that the 
DMR data shows slight evidence for non-Gaussianity, this has been attributed to
non-CMB anisotropy systematics.

Previous studies of the Gaussianity of the CMB anisotropy have been limited 
to large angular scales, in particular to the DMR maps. On these large angular 
scales the central limit theorem, in conjunction with sufficient smoothing,
ensures that any reasonable theoretical model will predict a Gaussian CMB
anisotropy. This does not hold on smaller angular scales. Therefore, it is 
important to investigate the Gaussianity of the CMB anisotropy on smaller 
angular scales.

Small-scale CMB anisotropy observations will soon be able to tightly constrain
cosmological parameters; see, e.g., Rocha (1999), Page (1999), and Gawiser
\& Silk (2000) for reviews. Current observational error bars are asymmetric 
(i.e., non-Gaussian), so accurate constraints on cosmological parameters can 
only be derived from a complete maximum likelihood analysis of current data
(see, e.g., Ganga et al. 1997; G\'orski et al. 1998; Ratra et al. 1999a;
Rocha et al. 1999) or good approximations thereof (Bond, Jaffe, \& Knox 2000), 
although even in this case one assumes that the underlying theoretical model is 
Gaussian, and it is important to explicitly test this assumption\footnote{
In fact, comparison of results from different experiments indicates evidence
for non-Gaussianity (Podariu et al. 2001), but this is almost certainly a
consequence of underestimated observational error bars rather than evidence 
for a non-Gaussian CMB anisotropy.}.
Derivation of tight constraints on cosmological parameters requires a combined
analysis of as many CMB anisotropy data sets as possible. For current data 
this requires a full maximum likelihood analysis (see, e.g., Ratra et al. 
1999b). If the CMB anisotropy is Gaussian and future observational error bars 
are symmetric (i.e., consistent with Gaussianity), then a $\chi^2$ comparison
of CMB anisotropy observations and model predictions will be a much faster way 
of determining constraints on cosmological parameters (see, e.g., Ganga, Ratra,
\& Sugiyama 1996; Dodelson \& Knox 2000; Tegmark \& Zaldarriaga 2000; Le Dour
et al. 2000).

In this paper we present results from a first test of the Gaussianity of the
small-scale (degree-scale) CMB anisotropy. We measure the genus statistic from 
CMB anisotropy maps of the QMAP and Saskatoon experiments (which have an angular resolution of about $1\deg$ FWHM). We also analyze a combination map of the 
data from these two experiments (QMASK, Xu et al. 2001). In the next section
we summarize the Wiener-filtering method used to construct noise-reduced CMB anisotropy maps and the method used to generate mock simulated survey maps.
In $\S$3 we compute the genus of the Wiener-filtered QMAP, Saskatoon, and 
QMASK maps. Our results are summarized and discussed in $\S$4 and we conclude
in $\S$5.

\section{Summary of Map-Making Processes}

\subsection{Wiener-filtered maps}

The QMAP (balloon-borne) and Saskatoon (ground-based) experiments were designed 
to measure the degree-scale CMB anisotropy in regions around the north 
celestial pole (NCP). QMAP measured in the Ka ($\sim 30$ GHz) and Q ($\sim 40$ 
GHz) bands, with six detectors in two polarizations (Ka1/2, Q1/2, Q3/4), 
with angular resolutions of $0\fdg89$, $0\fdg66$, and $0\fdg70$ FWHM for Ka1/2, 
Q1/2, and Q3/4, respectively. See Devlin et al. (1998), Herbig et al. (1998), 
and de Oliveira-Costa et al. (1998) for details of the experiment and the data 
from the two flights. The Saskatoon experiment measured the CMB anisotropy 
with an angular resolution of about $1\deg$ in the same frequency range as 
QMAP. See Netterfield et al. (1997) for details of the experiment and the 
data from three years of observations. Xu et al. (2001) have combined all QMAP 
data with Saskatoon data into a degree-scale CMB anisotropy map which they
call QMASK. This is the largest degree-scale CMB anisotropy map (648 square 
degrees). It has an angular resolution of $0\fdg68$ FWHM.  

The QMAP maps are very noisy, with large noise correlations between pixels.
The Saskatoon map is linear combinations of sky temperatures convolved with 
complicated synthesized beams. Wiener-filtering suppresses the noisiest modes 
in a map, thus emphasizing a statistically significant signal. Tegmark (1997)
reviews the general map-making process and discusses the various filters 
used. In what follows we summarize our Wiener-filtering of the QMAP data,
based on de Oliveira-Costa et al. (1998). We also apply this method to the 
QMASK data. The Saskatoon data are more complicated and for this we adopt the
Wiener-filtering method of Tegmark et al. (1997).

The QMAP data are composed of $n$-dimensional map vectors $X$'s
(the temperature at each of the $n$ pixels) and $n \times n$-dimensional 
covariance matrices $N$'s which characterize the pixel noise.  
The numbers of pixels of the QMAP data set are 2695 (1Ka1/2 band), 
2754 (1Q2), and 2937 (1Q3/4) for the 1st flight, and 1625 (2Ka1/2), 
1633 (2Q1/2), and 1591 (2Q3/4) for the 2nd flight. For the Saskatoon 
data set the number of data points (sky temperatures convolved with 
the synthesized beams) are 2586 for the NCP Cap data. The QMASK data, the 
combined version of QMAP and Saskatoon, has 6495 pixel temperatures.

For each experiment we compute the Wiener-filtered map $X_{\rm WF}$ given by
$$
   X_{\rm WF} = S[S+N]^{-1} X.      
   \eqno (1)
$$ 
Here $S$ is the CMB theoretical model covariance matrix defined as 
$$
   S_{ij} = \langle X_{\rm CMB} X_{\rm CMB}^T \rangle 
         = \sum_{\ell=2}^{\infty} {(2\ell+1) \over {4 \pi}} 
           P_\ell(\hat{r}_i \cdot \hat{r}_j) B_\ell^2 C_\ell,   
   \eqno (2)
$$
where we sum over multipole moments $\ell$, $X_{\rm CMB}$ is the CMB 
theoretical model temperature map vector, $P_\ell$'s are Legendre polynomials, 
and $B_\ell$'s are the spherical harmonic expansion coefficients of the beam of 
the experiment. For QMAP and Saskatoon, the CMB anisotropy theoretical spectrum 
we use is the flat bandpower spectrum $C_\ell = {{24\pi}\over 5} 
Q^2/\ell(\ell+1)$ normalized to $Q=30~\uK$. The global appearance of a map 
depends on the bandpower at the angular scales where the experiment is sensitive
(Tegmark et al. 1997). If we use a different power spectrum for the signal, 
e.g., a spatially-flat cosmological constant ($\Lambda$) dominated cold dark 
matter (CDM) model power spectrum, the temperatures of the filtered map change, 
but the overall distribution of hot and cold spots is conserved. The QMASK data 
contains more cosmological information than QMAP or Saskatoon alone, so we use
a flat-$\Lambda$ CDM model power spectrum (see $\S$2.2) to Wiener-filter 
QMASK. Such a spectrum more accurately summarizes available CMB anisotropy 
observations (e.g., Podariu et al. 2001).

The Wiener-filtered maps generally have the same angular resolution as the 
experimental beams. This is true for the QMAP and QMASK data. That is, these
maps have angular resolutions of $0\fdg89$, $0\fdg66$, and $0\fdg70$ FWHM for 
QMAP Ka1/2, Q1/2, and Q3/4, respectively, and $0\fdg68$ FWHM for QMASK. For 
Saskatoon, however, there is an additional $1\deg$ FWHM smoothing 
in the filtering process (see Tegmark et al. 1997 for details). 
Assuming a $1\deg$ FWHM angular resolution for the Saskatoon experiment, 
the total angular resolution of the Saskatoon Wiener-filtered map is 
approximately $1\fdg4$ FWHM.

To compute the genus we stereographically project the Wiener-filtered maps 
on to a plane. This projective mapping is conformal and locally preserves
shapes of structures. A Gaussian interpolation filter is used in the 
projection. We choose the radius of the Gaussian filter to be 2.5 times 
the pixel size, where the pixel sizes are those of the Wiener-filtered maps.
The total smoothing scales of the final maps are $1\fdg53$, $1\fdg6$, and $1\fdg42$ FWHM for QMAP, Saskatoon, and QMASK, respectively.

\subsection{Mock Survey Maps}

We have generated 50 Wiener-filtered mock survey maps for each experiment. To 
generate these CMB anisotropy maps we use a flat-$\Lambda$ CDM model spectrum 
with $\Lambda$ density parameter $\Omega_\Lambda = 0.6$, non-relativistic 
matter density parameter $\Omega_0 = 0.4$, $h = 0.6$ (where the Hubble constant 
$H_0 = 100 h$ km s$^{-1}$ Mpc$^{-1}$), and baryonic density parameter 
$\Omega_B = 0.0125 h^{-2}$ (Ratra et al. 1997). First the pure CMB anisotropy 
flat-$\Lambda$ model map is observed at 50 different parts of the sky using 
the method of Park et al. (1998). Then the instrumental noise is generated from 
the noise covariance matrix. The noise map vector of each experiment is 
computed from $\vec{n} = N^{1/2} \vec{n}_r$ where $N^{1/2}$ is the Cholesky 
decomposition of the noise covariance matrix and $\vec{n}_r$ is a vector of 
the same dimension as $\vec{n}$ and is composed of samples randomly 
drawn from a unit normal distribution (Netterfield et al. 1997, $\S$10).

The 50 pure CMB anisotropy signal map vectors are convolved with the beam 
of each experiment. They are then added to the instrumental noise map vectors
to produce the mock survey data. These CMB mock survey maps are Wiener-filtered 
and projected stereographically in the same way that the observational data are 
analyzed. The Wiener-filtered versions of the observed QMAP and Saskatoon CMB 
maps are shown in Figure 1, along with Wiener-filtered mock maps. 
The observed QMASK Wiener-filtered map is shown in the Figure 2$a$ together 
with mock QMASK maps (Fig. 2$c$ and 2$d$).

\section{Genus Analysis}

We use the two-dimensional genus statistic introduced by Gott et al. (1990)
as a quantitative measure of the topology of the CMB anisotropy. For the 
two-dimensional CMB anisotropy temperature field the genus is the number of 
hot spots minus the number of cold spots. Equivalently the genus at a 
temperature threshold level $\nu$ is
$$ 
    g(\nu) = {1 \over 2\pi} \int_C \kappa ds,  
    \eqno (3) 
$$
where $\kappa$ is the signed curvature of the iso-temperature contours $C$.
The genus curve as a function of the temperature threshold level 
has a characteristic S shape for a Gaussian random-phase field.

The theoretical genus per steradian of a two-dimensional Gaussian field
with correlation function $C(\theta)$ is (Gott et al. 1990)
$$ 
    g(\nu) = {1 \over (2\pi)^{3/2}} {C^{(2)} \over C^{(0)}} 
            \nu e^{-\nu^2/2}, 
    \eqno (4) 
$$ 
where $C^{(n)} \equiv (-1)^{n/2} (d^nC/d\theta^n)_{\theta=0}$, and the 
temperature threshold is $\nu(C^{(0)})^{1/2}=\nu\sigma$. For a CMB anisotropy 
temperature map convolved with a beam $B$ and a smoothing filter $F$, the 
genus per steradian can be expressed in terms of the CMB power spectrum 
$C_\ell$ as
$$ 
   g(\nu) = {1 \over 2(2\pi)^{3/2}} 
            {{\sum \ell(\ell+1)(2\ell+1)C_\ell B_\ell^2 F_\ell^2}
     \over  {\sum (2\ell+1)C_\ell B_\ell^2 F_\ell^2}}
            \nu e^{-\nu^2/2}.              
   \eqno (5) 
$$ 
The genus curve shape is fixed by the Gaussian random-phase nature of the 
anisotropy field and its amplitude depends only on the shape of $C_\ell$
and not on its amplitude.

Non-Gaussian features in the CMB anisotropy will affect the genus curve
in many different ways. Since the genus-threshold level relation is known 
for Gaussian fields, non-Gaussian behavior of a field can be detected from 
deviations of the genus curve from this relation (Park et al. 1998, 2001).
Non-Gaussianity can shift the observed genus curve to the left 
(toward negative thresholds) or right near the mean threshold level.
It can also alter the amplitudes of the genus curve at positive and negative
levels differently causing $|g(\nu=-1)| \neq |g(\nu=+1)|$.
We quantify these properties by using genus-related statistics.

First the shift of the genus curve $\Delta \nu$ (Park et al. 1992, 1998, 2001) 
is measured from the genus curve by minimizing the $\chi^2$ between the data 
and the fitting function
$$ 
    G = A_s \nu^\prime e^{-{\nu^\prime}^2 /2},     
    \eqno (6) 
$$
where $\nu^\prime = \nu - \Delta\nu$, and the amplitude $A_s$ of the genus
curve is allowed to have different values at negative and positive 
$\nu^\prime$. The fitting is performed over the range $-1.0 \le \nu \le 1.0$.
The second statistic is the asymmetry parameter which measures the difference 
in the amplitude of the genus curve at positive and negative thresholds (i.e., 
the difference between the numbers of hot and cold spots at positive and 
negative threshold levels, respectively). The asymmetry parameter is defined as 
$$ 
    \Delta g = A_H - A_C,	
    \eqno (7) 
$$
where
$$ 
    A_H = \int_{\nu_1}^{\nu_2} g_{\rm obs} d\nu 
         / \int_{\nu_1}^{\nu_2} g_{\rm fit} d\nu,	
    \eqno (8) 
$$
and likewise for $A_C$. The integration is limited to $-2.4 \le \nu \le -0.4$
for $A_C$ and to $0.4 \le \nu \le 2.4$ for $A_H$. The overall amplitude $A$ of 
the best-fit Gaussian genus curve $g_{\rm fit}$ is found from $\chi^2$-fitting 
over the range $-2.4 \le \nu \le 2.4$. Positive $\Delta g$ means that more hot 
spots are present than cold spots. In summary, for a given genus curve we 
measure the best-fit amplitude $A$, the shift parameter $\Delta\nu$, and the 
asymmetry parameter $\Delta g$.

We present the genus curves as a function of the area fraction threshold
level $\nu_A$.  $\nu_A$ is defined to be the temperature threshold level
at which the corresponding iso-temperature contour encloses a fraction
of the survey area equal to that at the temperature threshold level $\nu_A$ 
for a Gaussian field. The genus estimated at each threshold level 
is an average over three genus values with threshold levels $\nu_A$ shifted 
by 0 and $\pm 0.1$. The amplitude of the genus curves for each band and flight 
of QMAP are low because of the small survey area. So we have averaged all QMAP
genus curves over bands (Ka \& Q) and flights (1st \& 2nd) to get a total
QMAP genus curve. 

Figure 3 shows the observed genus curves (filled dots) of the Wiener-filtered 
versions of the observed QMAP and Saskatoon maps and their best fitting Gaussian
genus curves (solid curves). Open circles are the median genus curves from 50 
realizations of Wiener-filtered mock survey maps made using the Gaussian 
flat-$\Lambda$ CDM model. The error bars are the $\pm 68$\% uncertainty limits 
from the 50 genus values. The QMASK genus curve is shown in Figure 4$a$.
Table 1 lists the genus-related statistics $A$, $\Delta\nu$, and $\Delta g$ 
(only for QMASK) measured from each CMB map, as well as the results from the 
mock survey maps. The 68\% uncertainty limits are again estimated from the 50
$A$, $\Delta\nu$, and $\Delta g$ values obtained from the mock data.

\section{Discussion}

The observed genus curves of the QMAP and Saskatoon maps have shift parameters
consistent with zero, and are consistent with those of the simulated Gaussian 
model. Although there are some deviations in the observed genus curves from the 
Gaussian curves at high threshold levels, i.e., at $\nu_A \gtrsim 2.0$, for both QMAP and Saskatoon, these effects are probably not evidence for non-Gaussianity 
in the CMB anisotropy but are probably caused by the small areas of the observed fields. This is because the genus is forced to have non-zero values even at high threshold levels when it is given as a function of the area fraction threshold 
level.
 
The QMASK genus curve, however, has a different behavior. Its Gaussian fitting 
curve matches well the mock survey result and the shift parameter 
$\Delta\nu_{\rm obs} = 
+ 0.18$ is consistent with zero. From the 50 mock survey maps, the probability 
that the shift parameter $|\Delta\nu| \geq 0.18$ is 20\%. In the QMASK 
Wiener-filtered map there are many fewer hot spots near $\nu_A = +1$ than cold 
spots near $\nu_A=-1$. This difference in the number of hot and cold spots 
causes an asymmetry of the genus curve, $\Delta g_{\rm obs}=-0.37$. The 
phenomenological reason for this asymmetry is that the observed region of the 
first flight QMAP experiment in the QMASK map, especially the middle-left region
of Figure 2$a$, has many cold spots near the threshold level $\nu_A = -1$. We 
now consider a few possibilities which might cause this asymmetry.
      
The first possibility for the asymmetry in the genus curve is the statistical 
effect caused by a small survey area. From the 50 genus curves of the mock 
survey maps, we estimate the probability that a genus curve measured from 
the QMASK survey area has asymmetry parameter $|\Delta g| \ge 0.37$ to be 
18\%. Therefore, the asymmetry of QMASK genus curve is not a very significant 
effect. 

The second is the effect of Galactic foreground contamination in the QMASK 
data since the survey region is located at low Galactic latitude ($8\deg 
\lesssim b \lesssim 46\deg$). De Oliveira-Costa et al. (1997, 2000) have 
quantified the level of foreground contamination in the QMAP and Saskatoon data 
by using a cross-correlation technique. For the Saskatoon experiment they 
detect a marginal correlation with the Diffuse Infrared Background Experiment 
(DIRBE) maps, but no significant correlation with synchrotron emission maps.
They also find that the QMAP 30 GHz Ka-band data are significantly correlated 
with synchrotron maps but not with the DIRBE maps. However, an overall 
consideration of the effects of these foreground contaminations on the CMB 
temperature fluctuations show that the Saskatoon and QMAP data are not 
seriously contaminated by foreground sources since the foreground signal causes 
the CMB fluctuations to be overestimated by only a few percent. Because the 
foreground and CMB temperature fluctuations add in quadrature, a foreground 
signal with an amplitude a few tens of a percent of the CMB fluctuation will 
cause the CMB fluctuations to be overestimated by only a few percent.

In this study we investigate the effect of foreground contamination on genus 
appearance by
using the cross-correlation technique of de Oliveira-Costa et al. (2000)
to quantify the level of Galactic emission contamination in the QMASK data. 
The high resolution 100 $\mu \rm m$ dust map (Schlegel, Finkbeiner, \& Davis 
1998) and the 1420 MHz synchrotron map (Reich 1982; Reich \& Reich 1986) 
are used as foreground templates. The results, summarized in Table 2, show that 
the QMASK data are marginally correlated with Galactic foreground emission, 
especially with synchrotron emission. Here, the coefficients $\hat a$'s 
represent the correlation of foreground templates with the QMASK map and the
$\Delta T$'s are the corresponding temperature fluctuations ($\Delta T = 
\hat a \sigma_{\rm Gal}$, where $\sigma_{\rm Gal}$ is the standard deviation 
of the template map). The best-fit estimate of the foreground contribution to 
the QMASK map is obtained by summing the two foreground maps multiplied by the 
correlation coefficients $\hat a$. However, the genus curve measured from the 
foreground-subtracted QMASK map is also asymmetric (Fig. 4$b$ and Table 1). 
The asymmetry is somewhat larger ($\Delta g = -0.46$) but the probability that
the asymmetry parameter $|\Delta g| \geq 0.46$ is still only 10\%. The 
Wiener-filtered version of the foreground template used in this analysis is 
shown in Figure 2$b$. We have also computed the genus of the mock surveys 
after adding the Galactic foreground template to the mock survey CMB anisotropy 
maps. Adding the foreground causes negligible asymmetry and shift, compared 
with the results of the foreground-free mock survey. The genus curve estimated 
from the QMASK data with a Galactic cut at $b>20\deg$ also has essentially the 
same shape. Thus these foregrounds cannot be responsible for the degree-scale 
spots. It is most probable that the QMASK survey region has statistically more 
cold CMB spots and less hot spots.

\section{Conclusions}

We present results from a first test of the Gaussianity of degree-scale
CMB anisotropy. The observed QMAP, Saskatoon, and QMASK CMB anisotropy 
fields are Gaussian, given the errors. In combination with previous upper
limits on the non-Gaussianity of the large-scale (DMR-scale) CMB anisotropy,
these results indicate that the CMB anisotropy appears to be Gaussian all the
way from angular scales $>10\deg$ down to degree angular scales.   

Larger area CMB anisotropy maps with higher signal-to-noise will allow for a 
more confident test of Gaussianity of the degree and sub-degree scale CMB 
anisotropy. The recent BOOMERanG 1998 and MAXIMA-1 data appear promising for 
this purpose.

\acknowledgments

We acknowledge valuable discussions with S. Dodelson, L. Knox, P. Mukherjee,
L. Page (PI of the QMAP and Saskatoon experiments), T. Souradeep, and Y. Xu. 
CP and CGP acknowledge support from the BK21 program of the Korean Government 
and KOSEF. BR acknowledges support from NSF CAREER grant AST-9875031. MT 
acknowledges support from NSF grant AST-0071213, NASA grant NAG5-9194, and the 
University of Pennsylvania Research Foundation.

\clearpage

\begin{deluxetable}{crcccc}
\tablewidth{0pt}
\tablecaption{
Genus-Related Statistics Measured from the Wiener-Filtered Observed
QMAP, Saskatoon, and QMASK Maps
}
\tablehead{
\colhead{Wiener-Filtered Map} &
\colhead{$A$} &
\colhead{$\Delta\nu_{\rm obs}$} &
\colhead{$\Delta\nu_{\rm mock}$} &
\colhead{$\Delta g_{\rm obs}$} &
\colhead{$\Delta g_{\rm mock}$}
}
\startdata
QMAP         &  $2.51$ & $+0.11$ & $+0.00\pm0.10$ & $$  & $$ \\
Saskatoon    &  $6.58$ & $-0.04$ & $+0.00\pm0.11$ & $$  & $$ \\
QMASK        & $16.89$ & $+0.18$ & $+0.02\pm0.13$ & $-0.37$ & $-0.02\pm0.28$ \\
QMASK (Foreground Subtracted)
             & $16.78$ & $+0.16$ & $+0.02\pm0.13$ & $-0.46$ & $-0.02\pm0.28$ \\
\enddata
\end{deluxetable}

\begin{deluxetable}{ccccr}
\tablewidth{0pt}
\tablecaption{
Correlations between the QMASK Map and Galactic Emissions
}
\tablehead{
\colhead{Foreground Map} &
\colhead{$\hat a \pm \Delta \hat a$\tablenotemark{a}} &
\colhead{$\hat a / \Delta \hat a$} &
\colhead{$\sigma_{\rm Gal}$\tablenotemark{b}} &
\colhead{$\Delta T$ ($\mu \rm K$)\tablenotemark{c}} 
}
\startdata
100 $\mu\rm m$ dust  &  $1.88\pm4.17$ & $0.45$ & $3.78$  & $7.1\pm15.8$  \\
1420 MHz synchrotron &  $0.15\pm0.07$ & $2.01$ & $172.9$ & $25.9\pm12.1$  \\
\enddata
\tablenotetext{a}{$\hat a$ has the unit of $\mu\rm K (MJy/sr)^{-1}$
for the 100 $\mu\rm m$ template, and $\mu\rm K (mK)^{-1}$ for the 1420 MHz
template.}
\tablenotetext{b}{RMS levels in units of the foreground template maps. Units 
are MJy/sr for the 100 $\mu\rm m$ dust map, and mK for the 1420 MHz synchrotron
emission map.}
\tablenotetext{c}{$\Delta T \equiv (\hat a \pm \Delta\hat a )\sigma_{\rm Gal}$.}
\end{deluxetable}

\clearpage

\centerline{\bf FIGURE CAPTIONS}

\figcaption[]{Wiener-filtered maps for QMAP and Saskatoon CMB anisotropy
experiments. The top two rows show the Wiener-filtered observed QMAP maps 
for each band and flight. Note that the maps are not on the same scale; the
second flight maps are magnified by a factor of 2 relative to the first flight
maps. The next two rows show one set of the 50 Wiener-filtered mock QMAP 
survey maps. The Wiener-filtered observed Saskatoon map and two examples of
Wiener-filtered mock Saskatoon maps are shown in the bottom row. In all maps 
$\rm R.A. = 0\deg$ is at the top and increases clockwise.}

\figcaption[]{Wiener-filtered versions of $a)$ the observed QMASK data, $b)$ 
the Galactic foreground template, and $c)$, $d)$ two examples of mock QMASK
maps. The Galactic foreground shown in $b)$, subdominant to the CMB signal,
is amplified by a factor of five for ease of visualization. The dotted lines on 
the QMASK map in $a)$ indicate Galactic latitudes of $b=10\deg$, $20\deg$, and 
$30\deg$, from top to bottom.}

\figcaption[]{The genus curves (filled dots) measured from the Wiener-filtered 
observed maps of $a)$ the QMAP and $b)$ Saskatoon CMB anisotropy experiments. 
The solid lines are Gaussian curves, best fit to the measured genus points. 
The open circles are the median genus curve of 50 Wiener-filtered mock surveys 
in a Gaussian flat-$\Lambda$ CDM cosmological model.}

\figcaption[]{The genus curves (filled dots) measured from the Wiener-filtered 
observed maps of the $a)$ QMASK data and $b)$ QMASK data with known foreground 
contamination removed. The solid lines are Gaussian curves, best fit to the 
measured genus points. The open circles are the median genus curve of 50 
Wiener-filtered mock QMASK surveys in a Gaussian flat-$\Lambda$ CDM 
cosmological model.}

\clearpage

\begin{figure}
\vspace{-0.75truecm}
\begin{center}
\includegraphics[width=13cm]{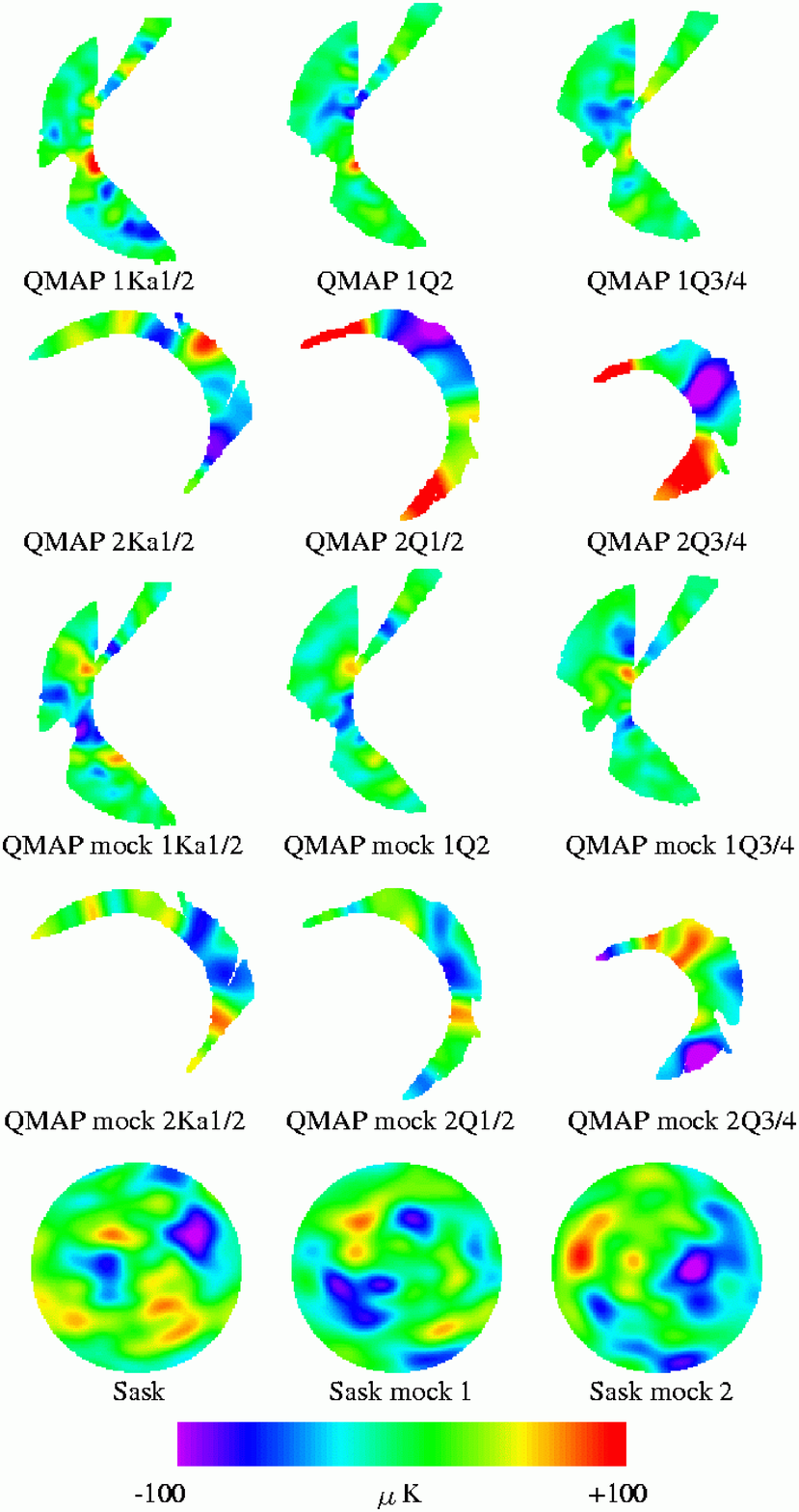}
\end{center}
Figure 1
\end{figure}
\clearpage

\begin{figure}
\resizebox{\textwidth}{!}{\includegraphics{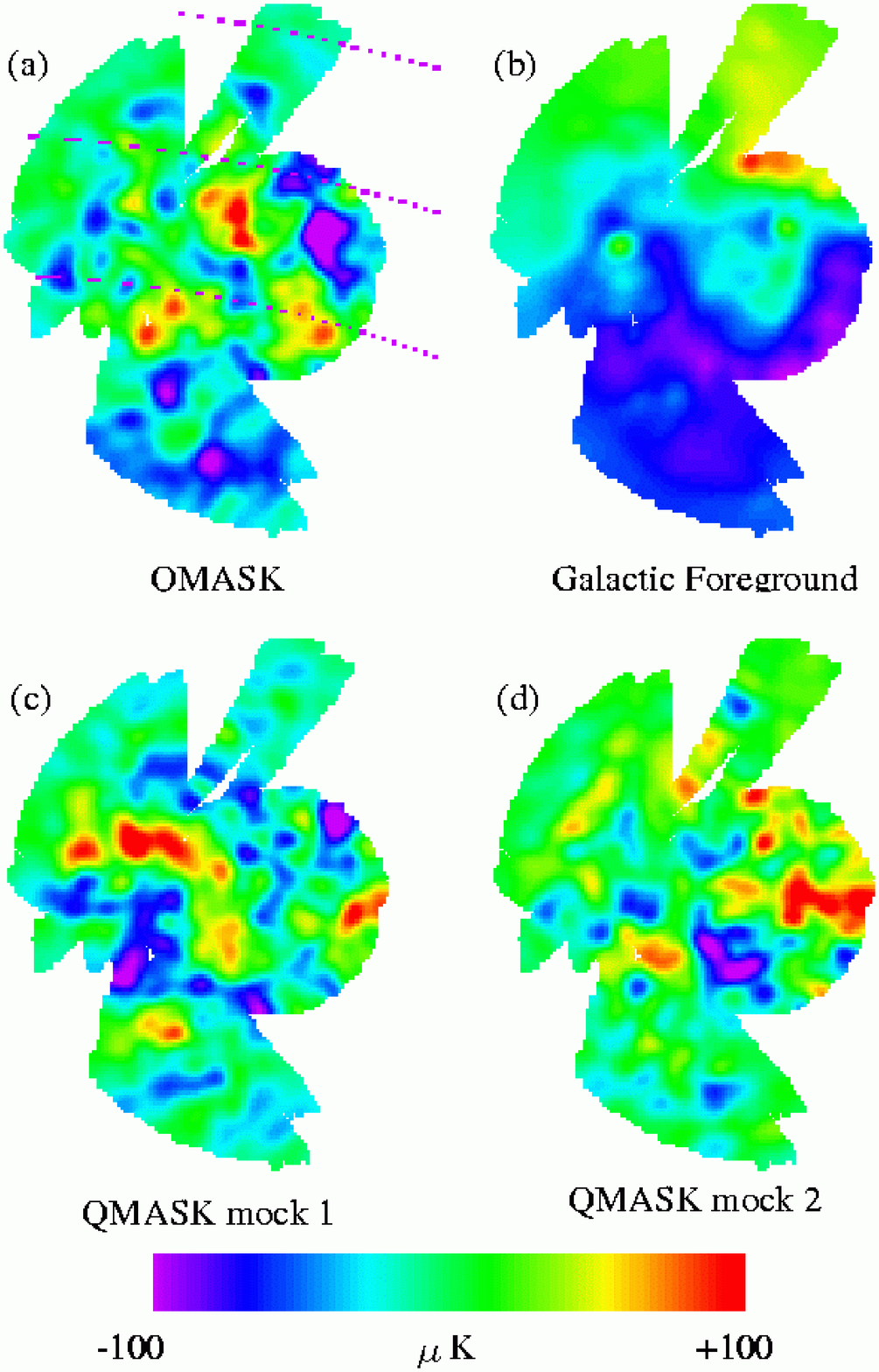}}
Figure 2
\end{figure}
\clearpage

\begin{figure}
\resizebox{\textwidth}{!}{\includegraphics{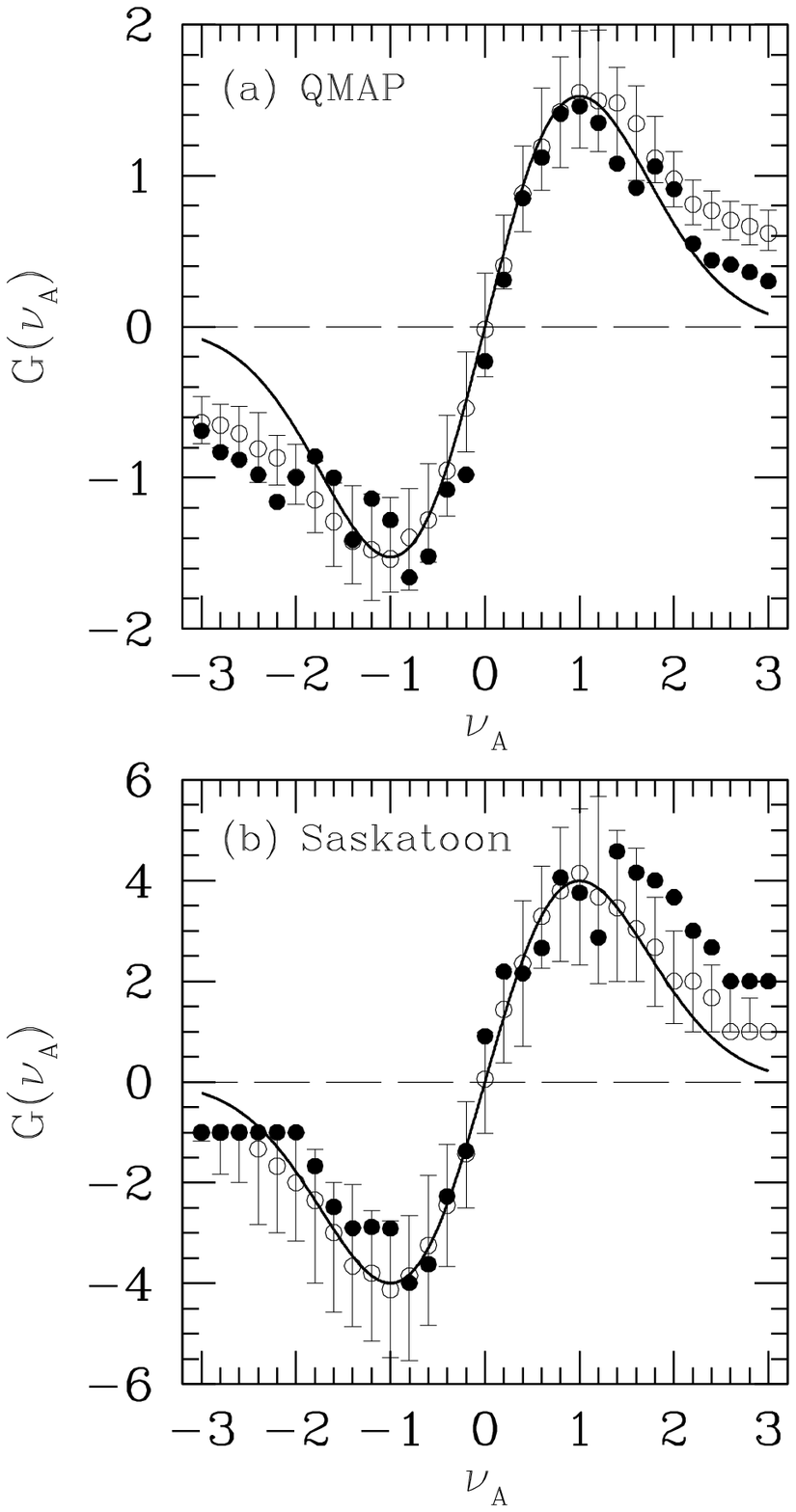}}
Figure 3
\end{figure}
\clearpage

\begin{figure}
\resizebox{\textwidth}{!}{\includegraphics{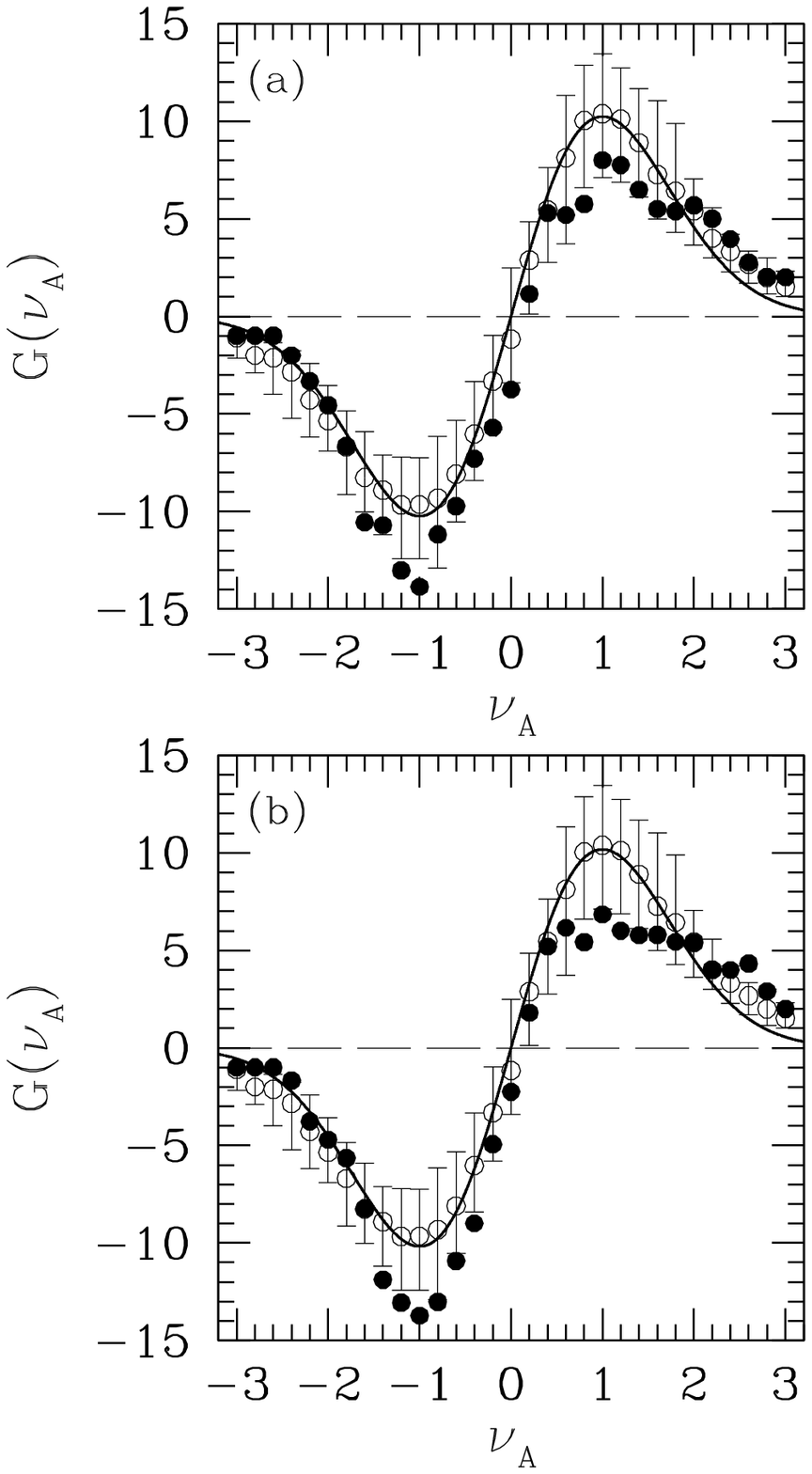}}
Figure 4
\end{figure}
\clearpage

\end{document}